# William Kruskal Remembered

**Stephen M. Stigler**

I knew Bill Kruskal as a dear friend and colleague for over 30 years, but I also knew him as a citizen of his department and university, a statesman of the statistics profession and a researcher in mathematical statistics. In all of those roles Bill showed characteristics he must have developed at an early age: unshakable integrity, consideration for others, painstaking attention to detail and an open, questioning scientific mind. In what I hope would be a spirit of social science inquiry that Bill would have sanctioned, I want to begin by asking a question of Bill that he asked so often of others.

For 30 years, whenever our department met in private session to face a decision on a tenure case, Bill would ask of his colleagues some version of this question: "Tell me," he would ask, "what specific significant new idea would you associate with the candidate; which of the candidate's works or publications are truly important?" Bill's purpose was clear—he wanted focus; he did not want to hear a recital of general impressions, he wanted evidence that would convince him, would convince the dean, would convince the provost and president. I will ask Bill's question about Bill himself, and advance some answers.

My first answer is that Bill will be remembered longest for a particular piece of research work during the 1950s. Bill was first appointed as an instructor in our newly formed department in 1950. The best known of his works is the Kruskal–Wallis test, a rank test for the analysis of variance he proposed in 1951 and then developed with Allen Wallis into a famous article published in 1952 (Kruskal and Wallis, 1952). This simple procedure has had a remarkable run. If you wish to know the extent of its fame, I suggest visiting *Google News*, as I did a few times shortly after Bill died. There the name Kruskal produced from 5 to 15 hits on the *Google News* pages (i.e., the search restricted to news sources of the past month) and almost all of those were to the use of the Kruskal–Wallis test in several different, newly released scientific studies. Indeed it is astonishing that a simple test proposed over a half century ago is still in current news. If a *Google News* count of 5–15 strikes you as meager, I suggest you try the same test on *Google News* using the name Gauss or Neyman or Pearson or Kolmogorov; in my trial all of these were either absent or merely single hits. If you try the full extent of *Google's* coverage, there are over 900,000 pages for Kruskal–Wallis. For this article alone Bill will be remembered as long as there are web pages, statistical software or textbooks.

If you protest (as Bill perhaps would) that the test is no more than a part of a proper analysis, and a small part at that, I say that misses my point. This longevity is significant evidence of Bill's marvelous ability to explain so clearly and develop his topic so thoroughly that in half a century no one has superceded him as a reference, in the manner that Robert K. Merton called "obliteration by incorporation." Bill's was the first word *and* the last word. Of course this was not his only major research success; he also made important contributions to the measurement of association, some with Leo Goodman, and to coordinate-free linear models and other areas. But Bill's question to his colleagues only asked for one idea: he wanted focus and the consequent detail.

I was careful in describing this test of Bill's as the work for which he will be longest remembered. I do not believe it was his most important contribution. To my mind Bill's greatest contribution was the furtherance of scientific collegiality in our department, in the University of Chicago, in the profession of statistics and indeed in the broad intellectual community of the nation. I would refer to the importance of his role in this as "inestimable," but I am

*Stephen M. Stigler is Ernest DeWitt Burton Distinguished Service Professor and Chairman, Department of Statistics and the College, and member of the Committee on Conceptual and Historical Studies of Science, University of Chicago, Chicago, Illinois 60637, USA e-mail: stigler@galton.uchicago.edu.*

sure Bill would protest because of course I am going to try to estimate it.

In all these spheres he was the soul of collegiality. He nurtured junior faculty. He helped students for whom he had no formal responsibility, with references and problem suggestions. He shared classroom examples and exam questions. By his example he taught us the importance and showed us the intellectual rewards of dedicated attention to teaching at all levels. He instilled in the members of our department both collegial mutual respect and a sense of integrity in furthering the mission of our university in ways that still guide us today and still set us apart from most statistics departments, indeed most academic departments in any discipline. Ours is still Bill's department.

A signal quality of Bill's was the way he built bridges between members of the faculty who shared interests in ways they might otherwise never have realized. When I prepared a manuscript, he would offer copious detailed suggestions and insist that I send it to a half dozen other people he was sure would be interested. When our present provost, a historian of ancient Rome, arrived as a new associate professor 20 years ago, it was Bill who sent me a copy of his paper on the use of anecdotes as data in ancient times. Bill's ability to forge links preceded *Google* by over 40 years and exceeded it in intellectual depth and the ability to recognize related ideas.

Now, not everyone who encountered Bill took immediately to his way of helping. Not every author of a ten-page double-spaced paper is grateful for ten single-spaced pages of typed comments, making suggestions, even gentle suggestions, ranging from allusions to work in areas you had never heard of, to grammar and spelling. At Bill's 70th birthday party 15 years ago, Fred Mosteller told the story of how he greeted such a very long letter from Bill by sitting down and starting to write an equally long letter back, explaining why in every instance he had done things the way he had. Jimmie Savage learned of this and wrote Fred a short note: "Dear Freddie, stop answering Bill's letter and fix the MS."

In going through some of Bill's papers I came upon a 1952 refereeing file that shows neatly how some people came to accept and even appreciate, however reluctantly, this capacity of Bill's. A paper had been submitted to *The Annals of Mathematical Statistics* by a well known West Coast statistician, and it was sent to Bill for review. Before the whole process was finished there were two more revisions and three referee's reports from Bill: first report, two pages; second report five pages, third report, eight pages. The following extracts from the author's replies to these reports tell a story:

> The referee... mistakenly believes I am working on a simpler problem....
> I am not willing to make any further alterations just to please a perfectionist referee. Life is too short to waste it on continual refinements of a paper.

And finally a letter from the associate editor:

> The author has just written that he now realizes that his confidence coefficient is in error and wishes to withdraw the paper to revise it. He is very apologetic about his rejection of your report earlier. With a less careful or observant job this error would have slipped through into print to everyone's embarrassment.

I have some sympathy for that author. Bill's perfectionism and insatiable intellectual curiosity taxed even him, as two unfinished and unpublished manuscripts he left on coordinate-free linear models and on chi-square statistics will attest. Happily, generations of Chicago students benefited from his courses on this material, and more recently our colleague Michael Wichura has taken the first of these subjects and developed it far beyond what Bill had done, into a fine textbook (Wichura, 2006). I worked with Bill for nearly two decades on an idea of his to explore the roles and appearances of the word "normal" in statistics. We did finally manage a modest paper (Kruskal and Stigler, 1997) on the topic, but it covered only a tiny fraction of the accumulated material Bill had unearthed—and that paper would have joined the other unpublished manuscripts if Bill had followed his preference and pursued several other avenues he suspected would be revealing.

Bill was a dear friend and trusted guide. He was a true gentleman; a man of firm opinions and strongly held values gently expressed, and he was open to other views and could abide all but the true scoundrels of life. His counsel will be missed, but his influence on me and on our department will long survive him.